\documentclass[aps,prl,nofootinbib,
twocolumn]{revtex4}
\usepackage{amssymb}
\usepackage{amsmath,color}
\usepackage{epsfig}
\usepackage{graphicx,epsf,epsfig}
\usepackage{bbm}
\usepackage{subfigure}
\usepackage{multirow}
\usepackage{setspace}
\usepackage{verbatim}

\begin{document}

\title{Comment on ``Is Dark Matter with Long-Range Interactions a Solution to \\ All
Small-Scale Problems of $\Lambda$CDM Cosmology?''}

\author{Bj\"{o}rn Ahlgren}
\email{bjornah@kth.se}

\author{Tommy Ohlsson}
\email{tohlsson@kth.se}

\author{Shun Zhou}
\email{shunzhou@kth.se}

\affiliation{Department of Theoretical Physics, KTH Royal Institute
of Technology, 106 91 Stockholm, Sweden}

\maketitle

In a recent Letter~\cite{Aarssen:2012fx}, van den Aarssen {\it et
al.}~suggested a general scenario for dark matter, where both the
dark-matter particle $\chi$ and ordinary neutrinos $\nu$ interact
with an MeV-mass vector boson $V$ via ${\cal L}
\supset g^{}_\chi \overline{\chi} \gamma^\mu \chi V^{}_\mu +
g^{}_\nu \overline{\nu} \gamma^\mu \nu V^{}_\mu$ with $g^{}_\chi$
and $g^{}_\nu$ being the corresponding coupling constants. Given a
vector-boson mass $0.05~{\rm MeV} \lesssim m^{}_V \lesssim 1~{\rm
MeV}$, one needs $10^{-5} \lesssim g^{}_\nu \lesssim 0.1$ to
solve all the small-scale structure problems in the scenario of
cold dark matter \cite{Aarssen:2012fx}.

Recently, Laha {\it et al.}~\cite{Laha:2013xua} found that the
$\nu$-$V$ interaction might lead to too large decay rates of
$K^- \to \mu^- + \overline{\nu}_\mu + V$ and $W^- \to l^- +
\overline{\nu}^{}_l + V$, indicating that the scenario proposed
in Ref.~\cite{Aarssen:2012fx} is severely constrained. However,
such experimental bounds can be evaded if the longitudinal
polarization state of $V$ is sterile or if $V$ is coupled to
sterile neutrinos rather than ordinary ones~\cite{Aarssen:2012fx,Laha:2013xua}.

\begin{figure}[ht]
\includegraphics[width=0.35\textwidth]{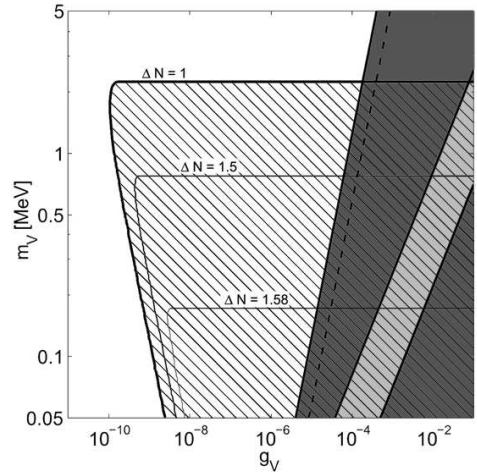}
\vspace{-1.5mm} \caption{Constraints on $g^{}_\nu$ and $m^{}_V$ from
$K$ decays (gray, solid line), $W$ decays (gray, dashed line), and BBN.
The two former are reproduced from Ref.~\cite{Laha:2013xua},
and the sample region in Fig.~3 of Ref.~\cite{Aarssen:2012fx} is also presented
(light gray). The hashed region bounded by the thick solid curve is excluded by
$\Delta N^{}_\nu < 1$ at $95\%$ C.L.~\cite{Mangano:2011ar}.
The excluded region will shrink for $\Delta N^{}_\nu < 1.5$ (see, e.g.,
Ref.~\cite{Hamann:2011ge}), and even further for the ``maximally conservative" limit
$\Delta N^{}_\nu \lesssim 1.58$ for $m_V > 0.05~{\rm MeV}$. \label{fig:limits}}
\end{figure}

Now, we show that the constraints on $g^{}_\nu$ and $m^{}_V$ from
Big Bang Nucleosynthesis (BBN) are very restrictive. In the early
Universe, $V$ can be thermalized via the inverse decay $\nu +
\overline{\nu} \to V$ and pair annihilation $\nu + \overline{\nu}
\to V + V$ and contribute to the energy density. If the
inverse-decay rate exceeds the expansion rate $H$ around the temperature
$T = 1~{\rm MeV}$, we obtain $g^{}_\nu > 1.5\times 10^{-10}~{\rm MeV}/m^{}_V$
for $m^{}_V \lesssim 1~{\rm MeV}$. For $m^{}_V \ll 1~{\rm MeV}$,
pair annihilation is more efficient than inverse decay to thermalize $V$.
Requiring the annihilation rate $\Gamma^{}_{\rm pair} \propto g^4_\nu T >  H$, one
obtains $g^{}_\nu > 3.4\times 10^{-5}$. For $m^{}_V > 1~{\rm MeV}$,
however, even if $V$ is in thermal equilibrium, its number density
will be suppressed by a Boltzmann factor.
To derive the BBN bound, we first solve the non-integrated Boltzmann equation for the
distribution function of $V$ for given $g^{}_\nu$ and $m^{}_V$, where
only the decay and inverse-decay processes are included in the computations
\cite{Kolb:1990vq, Kawasaki:1992kg, Basboll:2006yx, Garayoa:2009my, HahnWoernle:2009qn}.
Then, we calculate its energy density, and require the extra number
of neutrino species $\Delta N^{}_\nu < 1$ at $T = 1~{\rm MeV}$
\cite{Mangano:2011ar}. Thus, we can exclude a large region of
the parameter space, as shown in Fig.~\ref{fig:limits}. The contribution from
$V$ in thermal equilibrium reaches its maximum $\Delta N^{}_\nu \approx 1.71$ in
the relativistic limit.

Note that we have assumed $\Delta N^{}_\nu$ to be constant, but for
$m^{}_V > 1~{\rm MeV}$ it actually decreases during the BBN era, so
our constraint should be somehow relaxed in the large-mass region.
Since only the transverse polarizations of $V$ are involved in inverse
decay in the limit of zero neutrino masses, the BBN constraint does
not depend on whether the longitudinal polarization is thermalized or not.
If neutrinos are Dirac particles, the right-handed neutrinos $\nu^{}_{\rm R}$ can be
in thermal equilibrium as well. Both $V$ and $\nu^{}_{\rm R}$
contribute to the energy density, so one or more species of
$\nu^{}_{\rm R}$ is obviously ruled out. In addition, if $V$
is coupled to sterile neutrinos, which are supposed to be
thermalized, the BBN bound on $g^{}_\nu$ and $m^{}_V$ becomes more
stringent.

We are grateful to Torsten Bringmann for valuable communication.
Financial support from the Swedish Research Council and the
G{\"o}ran Gustafsson Foundation is acknowledged.

\end{document}